\documentclass[11pt]{article}
\usepackage{latexsym}
\usepackage[mathscr]{eucal}
\usepackage{amscd}
\usepackage{amssymb}
\usepackage{amsmath}
\usepackage{epsfig}
\usepackage{graphics}
\usepackage{enumerate}

\newcommand{\qed}{\hfill $\Box$}
\newcommand{\bN}{{\mathbb N}}

\newcommand{\R}{{\mathcal R}}
\newcommand{\cS}{{\mathcal S}}

\newcommand{\tc}{{\widetilde c}}
\newcommand{\tg}{{\widetilde g}}

\newcommand{\tO}{{\widetilde \Omega}}

\newcommand{\bT}{\mathcal T}
\newcommand{\scri}{\mathcal I}

\newcommand{\tal}{{\widetilde \alpha}}
\newcommand{\tbe}{\widetilde \beta}

\newcommand{\tC}{{\widetilde C}}

\newcommand{\lt}{\triangleleft}

\newcounter{def}

\newtheorem{lemma}{Lemma}
\newtheorem{proposition}{Proposition}
\newtheorem{corollary}{Corollary}
\newtheorem{claim}{Claim}

\newcommand{\bproof}{\setlength{\parindent}{0mm}{\bf Proof{~~}}}
\newcommand{\eproof}{$\Box$\setlength{\parindent}{5mm}}
\newcommand{\be}{\begin{equation}}
\newcommand{\ee}{\end{equation}}

\newcommand{\cP}{\Phi}
\newcommand{\stk}{a^1 \lt a^2 \lt \dots a^k}
\newcommand{\stl}{a^1 \lt a^2 \lt \dots a^l}
\newcommand{\stt}{a^1 \lt a^2}
\newcommand{\stinf}{a^1 \lt a^2 \lt \dots a^\infty }
\title{ Observables in Extended Percolation Models \\ of Causal Set Cosmology}
\author{Fay Dowker${}^1$ and  Sumati Surya${}^2$
\\
${}^1$Blackett Laboratory, Imperial College, London, SW7 2AZ, UK \\
${}^2$Raman Research Institute, Sadashivanagar, Bangalore, 560 080, India
}
\begin{document}
\maketitle
\begin{abstract}
Classical sequential growth models for causal sets provide an
important step towards the formulation of a quantum causal set
dynamics. The covariant observables in a class of these models known
as generalised percolation have been completely characterised in terms
of physically well-defined ``stem sets'' and yield an insight into the
nature of observables in quantum causal set cosmology.  We discuss a
recent extension of generalised percolation and show that the
characterisation of covariant observables in terms of stem sets is
also complete in this extension.
\end{abstract}
\section{Introduction}

In causal set quantum gravity, a classical stochastic analogue of a
quantum dynamics is provided by the classical sequential growth (CSG)
models \cite{Rideout:1999ub}.  Such stochastic models provide a useful
arena in which to investigate some of the thorny issues, such as
general covariance, with which we expect to have to grapple in the
quantum theory.  In these models, a labelled causal set grows, element
by element so that at each stage $n$, all possible transitions from an
$n$-element causal set to an $(n+1)$-element causal set are assigned
probabilities by the specific model.  If this growth  is (a)   Markovian,
i.e. obeys Markovian sum rules for the probabilities, and satisfies
(b) general covariance (GC) (label independence) and (c) Bell
Causality (BC) \cite{Rideout:1999ub} then the solution is a class of
models we call the Rideout Sorkin (RS) models.

It has been suggested that {\it quantum} sequential growth models can
be sought along similar lines, constructing a ``decoherence
functional'', rather than a probability measure, on the sample space
of causal sets, satisfying the appropriate quantum analogues of the
conditions (a), (b) and (c). However, as discussed in \cite{Henson:2004wb},
finding a quantum analogue of the Bell Causality condition  is a
difficult task and it is perhaps useful  to understand as much as
possible about the original condition in its CSG setting.

Bell Causality is given in \cite{Rideout:1999ub} as a condition on
ratios of transition probabilities, and is strictly defined only when
all transitions have non-zero probability; RS models satisfying this
condition are termed ``generic''.  In these models, any probability
$\tau$ for a transition at stage $n$ in which the new element has $\varpi$
ancestors, and $m$ parents, can be expressed in terms of a set of
coupling constants $\{ t_0 =1, t_1, \ldots, t_i, \ldots \}$:
\begin{equation}
\tau =  \frac{ \lambda(\varpi, m)}{\lambda(n,0)},
\end{equation}
where $\lambda(\varpi,m)=\sum_{i= m}^\varpi {\varpi-m \choose \varpi -i} 
t_i$. Each $t_i$ is not a transition probability, but the
ratio of the probabilities of the following two transitions: (i) the
``timid'' transition from the $i$-antichain (the causal set of $i$
unrelated points) in which the $(i+1)$th element is added to the
future of all elements in the $i$-antichain and (ii) the
``gregarious'' transition from the $i$-antichain to the
$(i+1)$-antichain. Denoting the probability of (ii) by $q_i$, 
$t_i$ can be expressed as \cite{Rideout:1999ub}
\begin{equation} \label{tis}
t_i = \sum_{k=0}^i (-1)^{i-k} {i \choose k} \frac{1}{q_k}.
\end{equation} 
If we specify the dynamics in terms of the free
parameters $\{t_i\}$, subject only to the conditions $t_i \geq 0$ and
$t_0=1$, certain transition probabilities can then vanish. 
Note 
however that $q_k$ cannot vanish for any $k\ge 1$ because 
\begin{equation}
q_k^{-1} = \sum_{i =0}^{k} {k \choose i} t_i\;.
\end{equation}

These models, defined by the $\{t_i\}$,
are called ``generalised percolation'' (GP) dynamics.  Thus, while a
GP is Markovian and generally covariant \cite{Rideout:1999ub} the BC
condition of \cite{Rideout:1999ub} is not well defined.

For these reasons, it is important to understand how the
vanishing of certain probabilities affects the principle of Bell
causality. In \cite{Rideout:1999ub}, Bell causality is defined as
\begin{equation} \label{bc}
\frac{ \alpha}{\widetilde \alpha}=\frac{\beta}{\widetilde \beta},
\end{equation}
where $\alpha, \beta $ are probabilities 
of two transitions from an $n$-element causal 
set $C_n$ with
a set of common ``spectators'' $S$. A spectator of a transition 
from a causal set $C_n$ is an element of $C_n$ which is not in 
the ancestor set of the newly born element.
$\widetilde \alpha, \,
\widetilde \beta$ are the probabilities of 
corresponding transitions from the
$(n-|S|)$-element causal set $C_n\backslash S$. 
For non-vanishing probabilities
this condition makes precise the physical requirement that
``spectators do not matter'', and that $\alpha$ and $\widetilde
\alpha$ are proportional to each other with a positive constant of
proportionality which is the same for the ratio of any other appropriately
defined pair of transition probabilities $\beta$ and $\widetilde \beta$.

In \cite{Rideout:2005} it is shown that an extension of
GP (which includes GP as a special case) 
is the general solution of GC and a weaker causality condition,
which allows transition amplitudes {\it including} the $q_k$ ($k>0$) to
vanish.  This condition, called Weak Bell Causality (WBC) is that (i)
condition (\ref{bc}) holds if all four transitions are non-vanishing; (ii) if
$\tilde \alpha=0$ then $\alpha=0$; (iii) if $\alpha =0 $ and $\beta
\neq 0$ then $\tilde \alpha =0$ and $\tilde \beta \neq 0$; (iv) If
$\alpha$ and $\beta$ are both zero, then nothing can be inferred about
the $\tilde \alpha$ and $\tilde \beta$.  The causets admitted by this
dynamics possess a characteristic feature: the existence of different
``eras'' of GP, each independent of the others.  We will refer to
this dynamics as ``extended percolation'' (EP) dynamics. 

The main focus of this work is to study the question of observables in
EP dynamics.  In \cite{Brightwell:2002vw} the set of covariant
observables is completely characterised for generalised percolation
dynamics in terms of physically comprehensible sets called ``stem
sets'': A stem $b$ in a causal set is analogous to a past set
$P=J^-(P)$ in continuum spacetime \cite{HawkingEllis}, {\it i.e.} $b$
is a stem if and only if $b=Past(b)$. A causal set $C$ is said to
contain a stem $b$ if there exist a labeling of $C$ such that the
first $|b|$ elements of $C$ are isomorphic to $b$. A stem set
$stem(b)$ is then the set of completed causets which contain $b$ as a
stem.  For any GP, the transition probabilities provide a measure on
the sample space of all labelled completed causal sets. This measure
can then be restricted to a covariant one on the space of all unlabelled
completed causets. The main result of \cite{Brightwell:2002vw} is that
in generalised percolation any measurable set is an element
of the sigma algebra generated by the stem sets, up to sets of measure
zero.

A characterisation of covariant observables in
terms of stem sets makes physical sense. In the continuum, for
example, a black hole in an asymptotically flat spacetime $(M,g)$ is
defined in terms of a past set: $J^-(\scri^+)\neq M$. The existence of
a black hole is clearly a covariant ``observable'' of the classical
theory. Another example of an observable is the number of bounces that
a given cosmology has undergone. What the result of \cite{Brightwell:2002vw}
shows is that asking stem questions will yield {\it all} the information
that  a GP can provide: the stem questions are physically complete. 
The hope is that a stem set characterisation of observables 
will be complete in quantum causal set dynamics as
well. In view of this larger goal, it is important to check the
robustness of this result for any EP dynamics, which
exhausts the class of Markovian dynamics compatible with general
covariance and WBC. The main aim of the
present work is to show that this indeed is the case.

In Section 2 we describe the key aspects of EP dynamics relevant to
the discussion of observables. In Section 3 we prove our main result,
namely that any covariant 
measurable set is an element of the sigma algebra generated
by the stem sets, up to sets of measure zero. In Section 4,
we discuss why Bell Causality needs to be framed as a condition on 
ratios  of transition probabilities.  We also consider an alternative 
causality condition we call Product Bell Causality (PBC)
--  $\alpha \widetilde
\beta= \widetilde \alpha \beta$ -- and show that it is strictly weaker
than WBC.  Both WBC and PBC therefore generalise (\ref{bc}) for the
case of vanishing transition probabilities.  In the appendix we show
that although PBC is strictly weaker than WBC, with the additional
requirement of general covariance, the resulting dynamics is the same.

\section{EP}


An EP model \cite{Rideout:2005} consists of a sequence of GP models run one
after the other, each for an ``{\sl era}'' consisting of some finite number
of stages. The resulting finite causets -- which we dub ``{\sl
turtles}" -- are placed one above the other in sequence to form a ``{\sl
stack}''. (To place a causet $a$ ``above'' causet $b$, set every element
of $a$ to the future of {\it every} element of $b$.)  Moreover, which
GP gets run at any era in the sequence, and how long it is to be run
for, depends on the realisation of the process -- {\it i.e.} the
actual causet generated -- in the eras preceding it. There may or may
not be a final infinite era in which the GP of that era is run to
infinity. If the latter does occur, the infinite causet generated in
that final era, which sits on top of a ``{\sl tower}'' of turtles, is called a
``{\sl yertle}'' \cite{Seuss}. 

We will elaborate on this description in what follows.

As in \cite{Brightwell:2002vw}
$\Omega(n)$ is the set of all $n$-element unlabelled causets, 
$\Omega(\bN)$ is the set of all finite unlabelled causets,  where $\bN$ is the
set of natural numbers and 
$\Omega$ is the set of completed (infinite) causal sets.  $\widetilde
\Omega(n)$, $\widetilde \Omega(\bN)$ and $\widetilde \Omega$ are their
labelled counterparts.  For any $\tc \in \tO$, let $\tc(n)$ denote the set of
the first $n$ elements of $\tc$.

We recall the useful terminology of ``break'': we say that a causet $c$ 
has a ``break at rank $n$ with past 
$c(n)$'' if it contains a stem, $c(n)$, of cardinality 
$n$ such that every element of the complement of $c(n)$ is above
every element of $c(n)$. A causal set
$c \in \Omega$ can have several breaks, but they must be ordered:
if the breaks occur at ranks $(n_1,n_2, \ldots n_k, \ldots )$ with
$n_i<n_{i+1}$, then $c(n_i) \subset c(n_{i+1})$.

\begin{claim}
If $c$ has a break at rank $n$ with
the past $c(n)$, then $c(n)$ is the unique stem
in $c$ with cardinality $n$. Equivalently, any natural 
labelling of $c$ must label $c(n)$ first. 
\end{claim}

\bproof 
Recall that a causet $c$ has a stem $b$ iff there is a natural 
labelling of $c$ in which $b$ is labelled first. 
Consider $e$ an element in the complement of $c(n)$. It is above every 
element of $c(n)$ and in any natural labelling, its label must be 
greater than $n$. Therefore $e$ cannot be 
contained in a stem of cardinality $n$. 
\eproof

Examples of breaks: An infinite chain has an infinite number of breaks
where the pasts of these breaks are the set of finite chains.  Another
example of causets with breaks are those formed in originary dynamics: 
all causets generated by this dynamics have a break at rank 1 with
the past of the break being the single element causet.

Let $a^i \in \Omega(n_i)$, $i=\{1,2, ... k\}$ where $n_i \in \bN$ with
$k \in \bN \cup \{ \infty \}$. The {\sl stack} $\stk$ is formed by putting
$a^2$ above $a^1$, $a^3$ above $a^2$ {\it etc}.  It is an $\sum_i n_i$
element causet with breaks at ranks $\sum_{i=1}^l n_i$, $l \in \{1,
\ldots k \}$ with pasts $\stl$. Note that a stack may have other
breaks but the ones at $\sum_{i=1}^l n_i$, $l \in \{1, \ldots k \}$
are specified as part of the definition of the stack. The causets
between the specified breaks in a stack are finite by definition. We
will say that a causet $c\in \Omega$ contains the stack $\stk$ 
if $c$ has a break at rank $\sum_in_i$ with past $\stk$.

This is a useful concept because an EP generates a causet with
breaks between each of the turtles.  The notion of breaks and stacks
can be unambiguously extended to labelled causets since they are 
label invariant concepts. We can think of the
dynamics during each era as a GP ``relative'' to the causet which has
already occurred in the previous eras.
Let $\tc_n \in \tO(n)$ have a break at stage $m<n$ with past
$\tc(m)$. Consider a transition from $\tc_n \rightarrow \tc_{n+1}$
such that $\tc_{n+1}$ also has a break at rank $m$ with past
$\tc(m)$. We will refer to a transition which ``preserves the break''
as a {\sl transition relative to $\tc(m)$}.  $c \in \Omega(|a|+k)$ is
an {\sl $a$-relative $k$-antichain} with a break at rank $|a|$ and
past $a$ if  $c\backslash a$ is a $k$-antichain, i.e. it is  a
$k$-antichain stacked on top of $a$. Let $q_{(a)\,i}$ be the
probability for the transition from the $a$-relative $i$-antichain to
the $a$-relative $(i+1)$-antichain for $i>0$, and $q_{(a)\,0}$ the
probability of the 
timid transition from $a$.  As in (\ref{tis}) we define the {\sl
coupling constants relative to $a$}:
\begin{equation}
t_{(a)\,i} = \sum_{k=0}^i (-1)^{i-k} {i \choose k} \frac{1}{q_{(a)\,k}},
\end{equation}
which is the ratio of the following two transition 
probabilities: (i) the probability of the timid
transition from the $a$-relative $i$-antichain and (ii) $q_{(a)\,i}$.

Since an EP is specified iteratively, era by era,
it will be useful to introduce  a compact notation for the coupling
constants for any era,  showing 
the dependence on the causet generated in previous eras. 
After the $k$th era of an EP, the resulting causet will be a 
stack, $a\equiv \stk$. The $(k+1)$th 
era is given by a number  $n(a) \in \bN \cup \{\infty\}$ and 
a set of $n(a)$ coupling constants
relative to $a$,   $\{t_{(a)\,0}= 1, t_{(a)\,1} \ldots
t_{(a)\,n(a)-1}\}$.  We will use the
notation $\cP(a,n(a))$ to denote these coupling constants. 
\footnote{Note that the couplings for the $k+1$th era are not only
relative to $a$ but also depend on $a$. To be strict we'd need to add
a further explicit dependence on $a$ to the $t$'s. However, this would
encumber the already heavy notation and we retain the current form for
simplicity.}  We can construct each era explicitly as follows:

\noindent {\sl First Era: } In this first era, one starts with the empty
set $\emptyset$, and the dynamics is given by $\cP(\emptyset,
n(\emptyset))$, $n(\emptyset) \in \bN \cup \{ \infty\}$. 
This dynamics is equivalent to a probability distribution on 
$\Omega(n(\emptyset))$. 

If $n(\emptyset) = \infty$ this is the whole EP: it is a single
GP. 

If $n(\emptyset) < \infty$, we define 
 $E(\emptyset) \subseteq \Omega(n(\emptyset))$ to
be the set of causets with non-zero probability,
with the equality holding only if
none of the transition probabilities up to stage $n(\emptyset) -1$
vanish.

\noindent {\sl  Second Era:} For each
element, $a^1$ of $E(\emptyset)$,
there is an $a^1$-relative 
GP $\cP(a^1, n(a^1))$, which generates a probability 
distribution on the set of $(n(\emptyset) + n(a^1))$-element 
causets which have a break at rank $n(\emptyset)$ with past 
$a^1$. 

If $n(a^1) = \infty$ the process $\cP(a^1, n(a^1))$ is run to 
infinity. 

If $n(a^1) < \infty$ define $E(a^1)\subset \Omega(n(\emptyset) +n(a^1))$
to be the set of causets with non-zero probability. 
The inclusion is strict, since the growth
only allows causets which have a break at rank $n(\emptyset)$ with
past $a^1$. 

Notice that different
$a^1 \in E(\emptyset)$ give different second-era dynamics that are
independent of each other, hence the need for the labeling
$(a^1)$.

\noindent{\sl Third Era:} For each $\stt \in E(a^1)$ 
there is an $(\stt)$-relative 
GP $\cP(\stt, n(\stt))$. If $n(a^2) < \infty$, let
$E(a^1\lt a^2) \subset \Omega(n(\emptyset) + n(a^1) + n(a^2))$
be the set of causets with non-zero probability.
 And so on. 

A turtle is the set of all elements born during a single
finite era.  A yertle is the set of all elements born during an
infinite era (necessarily the last).

Define $\Xi(\bN)$ to be the set of finite stacks 
with non-zero probability in the EP, i.e.,  
\begin{equation} 
\Xi(\bN) = \bigcup_k E(\stk)\ . 
\end{equation} 

Define $\Xi(\infty)$ to be the set of infinite stacks $\stinf$ such that all
their finite substacks are in $\Xi$. 
Define $\Xi \equiv \Xi(\bN) \cup \Xi(\infty) \cup
\{\emptyset\}$. We will say that a stack is {\it admitted} by the
dynamics if it is an element of $\Xi$. 

 \underline {Special Cases:}

\begin{enumerate} 

\item Generalised percolation is a special case of EP with 
a single era: $\Xi
  =\{\emptyset\}$, $n(\emptyset) = \infty$. 

\item An ``originary'' generalised percolation is specified by 
taking a GP and putting the causet generated above a single 
minimal element (the ``origin''). Since $q_{(\emptyset)\,1}=q_1=0$ it is
not a GP, but an EP with $n(\emptyset)=1$ in the first 
era and $n(\{.\}) = \infty$ in the second, 
where $\{.\}$ is the single element causal set. 
\end{enumerate}

\section{Main Result}

Following \cite{Brightwell:2002vw}, for each EP we have $\Omega$ 
the sample space of all unlabelled, past finite, completed causets,
$\R$ the collection of physical measurable sets and $\R(\cS)$
the sigma algebra generated by $\cS$ the family of all stem 
sets. The identification of $\R(\cS)$ as the complete set of 
physical covariant questions 
for GP is the main result of \cite{Brightwell:2002vw}.  We want to
prove this same result for EP:  

\begin{proposition}
In an EP, the family of stem sets, $\cS$, generates the sigma algebra,
$\R$, of covariant measurable sets 
up to sets of measure zero.
\end{proposition}

As in \cite{Brightwell:2002vw} a crucial concept is that of a rogue
which is a causet which has at least one ``clone,'' where a clone of a causet
$c$ is a non-isomorphic causet which has the same stems as $c$. The
set of all rogues is written $\Theta$. 
An important kinematical result of \cite{Brightwell:2002vw} is a
characterisation of the set of rogues $\Theta$: $c \in \Theta$ iff $c$
contains a level with infinitely many non-maximal 
elements 

An example of an infinite level is the first (and only level) of an
infinite antichain. However, all the elements in this level are
maximal and hence the infinite antichain is not a rogue causal set. It
is shown in \cite{Brightwell:2002vw} that $\Theta$ can be built by
performing countable set operations on stem sets so $\Theta \in
\R(S)$. Since rogues are not specified by their stems, their occurrence
would be an obstacle to proving Proposition 1.  We need to understand
how rogues can arise in EP dynamics.

We define $\bT$, the set of {\sl towers}  
\begin{equation} \label{tower}
\bT = \{ x\in \Xi: x \subset y, \ y\in \Xi \implies 
x = y\}\,.
\end{equation}
$\bT$ contains all the infinite 
admitted stacks and those finite admitted stacks such that the
next era is infinite. 
We first notice that

\begin{lemma}\label{mainlemma}
An admitted  stack  
$\stk\in \Xi$ 
is a stem of a causet $c$ generated by
the dynamics iff $c$ has a
break with past $\stk$. 
Moreover, any causet with this stack 
as a stem can be generated by only one sequence of GPs 
for the first $k$ eras  and the $(k+1)$th era will be 
 the GP $\cP(\stk ,n(\stk))$.
\end{lemma}
\bproof $\stk \in \Xi$ and so there's a sequence
of GPs that generate it: $\{\cP(\stl), n(\stl) \}$, $l = 0, 1, \ldots
k$, with $a^0=\emptyset$. 

Assume $\stk$ is a stem of a causet, 
$c$, generated by the EP. The initial era is fixed to be the GP 
$\cP(\emptyset, n(\emptyset))$. Since
$| a_1 | = n(\emptyset)$, either all elements of $a_1$ are born in the
first era, or there is some element $x \notin a_1$ which is born in
the first era and some $y \in a_1$ born in a subsequent era. Since 
this is an EP, this implies $x\prec y$, which
means $a$ is not a stem which is a 
contradiction. Therefore $a^1$ is generated in the first era and the second
era must be $\cP(a^1,n(a^1))$. $| a^2| = n(a^1)$ and either all elements
of $a^2$ are born in the second era or not. The latter possibility
leads to a contradiction as before, so that the third era is
$\cP(\stt,n(\stt))$. And so on through the $k$th era. 
So $c$ has a break with past $\stk$. 

Therefore if  $\stk$ is a
stem of $c$, then it is generated by the GPs $\{ \cP(\stl,
n(\stl)) \}$, $l \in \{ 0, 1, \ldots k\}$. 

Assume that $\stk$ is a stack in $c$, i.e.  $c$ has a break with past $\stk$
Then by Claim 1, it is a stem in $c$.  
\eproof

\begin{corollary} 
Let $\tau, \tau'
\in \bT$.  If $\tau$ and $\tau'$ are both towers in 
the same causet $c$ generated by
the dynamics then $\tau = \tau'$. 
\end{corollary}

It is therefore possible to divide the set of causets that can occur into
disjoint sets based on a classification of complete towers. 
Since no tower contains an infinite level, causets which are infinite 
towers cannot be rogues. A causal set  that is generated by the 
dynamics and contains a finite 
tower $\tau \in \bT$
will have a yertle stacked on top of the tower and will be a 
rogue iff the yertle is a rogue.
Any finite $\tau
\in \bT$ can be classified according to the GP $\cP(\tau, n(\tau)= \infty)$:

\begin{enumerate}[(a)]

\item \label{dust} $t_{(\tau)\,0} =1$, $t_{(\tau) \,i} = 0, \forall
i > 0$. This is a deterministic ``dust dynamics'' which almost surely
produces a yertle which is the infinite antichain and hence not a
rogue.
 
\item \label{forest} $t_{(\tau)\,0} =1$,
$ t_{(\tau)\,1} \ne 0$, $t_{(\tau)\,i} = 0,
\forall i >1$. This is also deterministic, the ``Forest dynamics'',
and almost surely produces the ``Forest yertle'' which consists of
infinitely many trees in which every element has infinitely many
descendants and every element, except the minimal elements, has exactly
one ancestor.  The Forest yertle is a rogue. 

\item \label{regular}$t_{(\tau)\, 0}= 1$, $t_{(\tau)\,1} \ne 0$ and
$t_{(\tau)\,i} \ne 0$ for some $i \ge 2$. It was proved in 
\cite{Brightwell:2002vw} that such dynamics cannot  produce
rogues.

\end{enumerate} 

For a pure GP dynamics, the Forest dynamics is the only one that can
produce a rogue, but it is deterministic and that is enough to prove
that that the stem sets generate the sigma algebra of covariant
measurable sets up to sets of measure zero.  
An EP is made of pieces of GP's and
so all three types can occur in a single EP. 
Although a Forest yertle dynamics is deterministic, an EP dynamics 
which contains it may not be and one requires a proof different from that
of \cite{Brightwell:2002vw}.  We will henceforth refer to causets with
yertles of type (\ref{forest}) as {\sl forested towers.}  A clone of
a forested tower $c$ is a causet non-isomorphic to $c$ which has the
same stems as $c$. If $\tau$ is the tower in $c$, then every clone of
$c$ is a causal set with a tower $\tau$ with a clone of the Forest
on top of it.

We are now in a position to prove our main theorem, restated in the
following way.

\begin{proposition}\label{main.theorem}
In any EP, for every set $A\in\R$, there is a set $ B\in \R(\cS)$ such
that $\mu(A\triangle B) =0$.
\end{proposition} 

\bproof 

Consider a measurable set $A \in \R$. 
Consider the set, $T$, of finite towers corresponding to 
all the forested towers in $A$. $T$ is countable so we 
can list the elements of $T = \{\tau_1, \tau_2, \dots\}$.
Since each 
$\tau_k$ is a tower,  
 lemma \ref{mainlemma} implies that $n(\tau_k)= \infty$ 
and $\cP(\tau_k, n(\tau_k))$ is the Forest dynamics.  

Define 
\be
F  \equiv  \bigcup_i stem(\tau_i)
\ee
and let $B = A\cap \Theta^c \sqcup \Theta \cap F$. 
In \cite{Brightwell:2002vw} it was proved that $A\cap \Theta^c$ is
an element of $\R(\cS)$ and the proof is independent of
the measure and so holds here also. $F \in \R(\cS)$ and so 
$B \in \R(\cS)$ also. 

We have 
\be
B\Delta A = A\cap \Theta \cap F^c \sqcup A^c \cap \Theta \cap F 
\ee
and
\be
\mu(B\Delta A) = \mu(A\cap \Theta \cap F^c) + \mu (A^c\cap \Theta \cap F). 
\ee
$A\cap \Theta \cap F^c$ contains only rogues in A 
which do not have any $\tau_k$ as
a stem. So these rogues are not forested towers. 
Rogues which are not forested towers 
almost surely do not happen (proof below). $A^c \cap\Theta \cap F$ 
is the set of rogues, not in $A$, which have at least one $\tau_k$ 
as a stem. By lemma \ref{mainlemma}, a causet $c$ generated by the
dynamics which has $\tau_k$ 
as a stem must have a break with past $\tau_k$. And the subsequent 
GP must be the Forest dynamics. Since an element of $A^c \cap\Theta \cap F$
cannot be a tower $\tau_k$ with the Forest above (because that 
is an element of $A$) it must be a tower $\tau_k$ with a 
clone of the Forest above, and these almost surely do not occur 
(see below).

So $\mu(B \Delta A) = 0$.

\eproof

\begin{claim}: Rogues that are not forested towers almost surely do 
 not happen. 
\end{claim}
\bproof
Rogues must have a level
containing infinitely many non-maximal elements. The
only  place an infinite level can occur in a causet 
generated by an EP is in a
yertle.  
The only yertle dynamics which can generate a causet with 
a level with infinitely many non-maximal elements is the 
Forest dynamics which almost surely generates the Forest. 
\eproof

\section{Generalisations of the Bell Causality Condition}

\subsection{Why Bell Causality involves ratios}

Let us reflect on the reason that the alternative condition $\alpha =
\widetilde\alpha$ is not the appropriate one for Bell causality. Let
us call this condition Very Strong Bell Causality (VSBC). In CSG
models there is no background causal structure, rather the events
themselves are the (growth of the) causal structure.  When there {\it
is} a background causal structure, the two conditions BC and VSBC (or
rather their natural analogues) are equivalent, but in CSG models
where there's no background, and where the conditions are expressed in
terms of the unphysical parameter time labelling stages, 
they are very different
as explained below.

VSBC is extremely strong in the
context of CSG, indeed it would imply that the only possible
dynamics are the one in which, almost surely, the
infinite antichain grows and the one in which, almost 
surely, the causal set which is the infinite antichain 
above a single minimal element grows.

First, to see that there is a problem consider all possible
transitions from a finite
causal set $B$, $B \rightarrow B_i, i = 1,2, \dots
n$.  The associated transition probabilities $P(B \rightarrow B_i)$
sum to 1. Now, consider a causal set $C$ which contains $B$ as a
stem. For each transition $B \rightarrow B_i$ there is a corresponding
transition $C \rightarrow C_i$ for which all the elements in $C$ that
are not in $B$ are spectators. Therefore, we must have $\sum_i
P(C\rightarrow C_i) = 1$ also.  But these will not exhaust the
possible transitions from $C$, and this condition will then force all
the other ones to have zero probability.

Indeed, starting with the transition from a single element to the
two-chain (with prob $p$) or the two-antichain (with prob $1-p$), you can
quickly show that you run into contradictions unless $p = 1$ or $0$. In
figure 1, 
\begin{figure}[ht]
\centering
\resizebox{2.5in}{!}
{\includegraphics{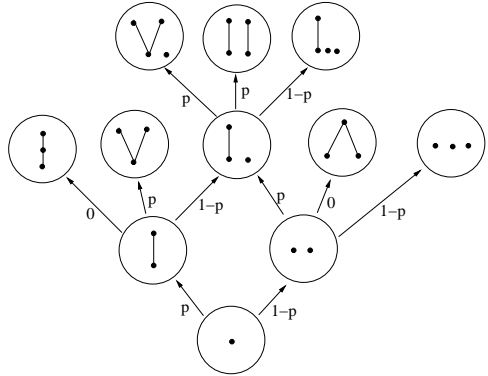}}
\caption{\small{Transition probabilities for VSBC}
\label{poscau.fig}}
\end{figure}
some transitions are shown with probabilities deduced from
VSBC. Consider the possible transitions from the 3-element causet
``L'', which is the union of a two chain and a single unrelated
element. Three are shown, two of which have probability $p$ and one
has probability $1-p$. This implies that either $p =0$ or the ``L''
has zero probability which can only happen if $p=0$ or $p=1$. If $p=0$
then only the infinite antichain can be produced.  If $p =1$ then the
``infinite antichain over one minimal element'' is produced.

The reason that VSBC is so restrictive here is that the causal 
structure is itself dynamical. 
In the context of a theory with a fixed background causal structure the
two conditions VSBC and BC (with the transition probabilities interpreted as
conditional probabilities for events occurring in specified
spacetime regions) are actually equivalent because the events
one is considering occur in {\it fixed} spacelike separated regions
$A$ and $B$, say.  Let $B$ be the region of the spectator events, then
one can enumerate an exhaustive set of events in region $A$. The
conditional probabilities of these sum to one (something must
occur in region $A$) and so equality of all their ratios implies
equality of the conditional probabilities themselves.

In CSG there is no fixed causal structure and the Bell causality
condition is imposed on the dynamics at the level of the
{\it labelled} (non-covariant) process.   In
the labelled process, the transition probabilities are for the
``next birth'' but there's no reason that the next element has to
be born ``in region $A$" -- and if there are very many
spectators around, it is very likely not to be.  This
"draining of probability from region $A$ births" conflicts with
VSBC, and is the underlying reason why VSBC
forces a lot of transition probabilities to be zero -- too
many, as pointed out above, leaving the dynamics trivial.

Condition BC however remains ``ok'', and we believe it
retains a large degree of plausibility.  It has the good
property that its closest analog in the standard (background
causal structure) situation is equivalent to the usual causality
condition (the one that gives rise to the Bell inequalities 
and goes by a variety of names, see \cite{Henson:2004wb}).  
And it is very restrictive while still allowing an
interesting family of dynamics.  Whether or not it is {\it the}
unique condition that one can legitimately identify as
{\it physical} Bell causality remains unsettled. Whether, in
particular, its implications for the covariant measure can be
discovered/understood in covariant terms -- in terms of stem
predicates -- is an interesting open question.

\subsection{ PBC is weaker than WBC}

Product Bell Causality (PBC) satisfies conditions (i) and (iv) of WBC,
but not conditions (ii) and (iii). Namely, if $\tal=0$, then either
$\alpha,$, or $ \tbe$ or both must vanish. If $\alpha \neq 0$ and
$\tbe =0$, then this violates condition (ii) of WBC. Condition (iii)
is then similarly violated: If $\alpha=0,$ $\beta \neq 0$ then while
this implies $\tal =0$, it does not necessarily imply the
non-vanishing of $\tbe$ (required by WBC for consistency with (ii)).

\begin{figure}[ht]
\centering
\resizebox{2.0in}{!}
{\includegraphics{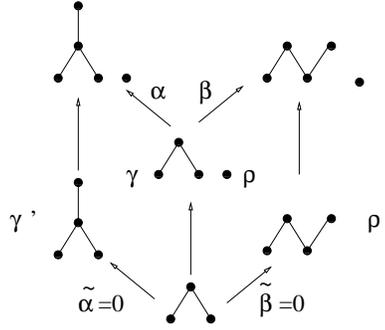}}
\caption{\small{Example of $\tal=0$, $\alpha \neq 0$, illustrating  the
    difference between WBC and PBC.}
\label{pbc.fig}}
\end{figure}

We show that violations of conditions (ii) and (iii) are compatible
with PBC. Let us consider the transitions $\alpha: C_{n-1} \rightarrow
C_n$, $\beta: C_{n-1} \rightarrow C_n'$, with a non-empty set of
common spectators $S$. Let $\tC_{m-1}=C_{n-1}\backslash S$,
$m=n-|S|$. Let $\tal: \tC_{m-1} \rightarrow \tC_m$, $\tbe: \tC_{m-1}
\rightarrow \tC_m'$ be the corresponding transitions without
$S$. Define the following sets of paths in the space of causets
$\gamma \equiv(C_1, C_2 \ldots C_n )$, $\gamma' \equiv(C_1, \tC_2
\ldots \tC_{m-1}, \tC_m, \ldots C_n )$, $\rho \equiv(C_1, C_2' \ldots
C_{n-1}, C_n' )$, $\rho' \equiv(C_1, \tC_2' \ldots \tC_{m-1}, \tC_m',
\ldots C_n' )$, where $C_1$ is the one element causal set.  While
$\gamma$ and $\gamma'$ intersect at $C_1$ and at $C_n$, we also
require that they do not intersect at $\tC_m$. One can always find
such $\gamma, \gamma'$ in the space of causets.  Similarly, $\rho$ and
$ \rho'$ are required not to intersect at $\tC_m'$. We show an example
in Fig. (\ref{pbc.fig}).
 \vspace{0.5cm}
\vspace{0.5cm}
Now, if $\tal =0$,  $Prob_{\gamma'}(\tC_m)=0$ and therefore
$Prob_{\gamma'}(C_n) =0$. If $\alpha \neq 0$, then PBC requires that
$\tbe=0$, so that $Prob_{\rho'}(\tC_m')=0$ and hence
$Prob_{\rho'}(C_n')=0$. However, if only PBC is imposed, it is
possible for $Prob_{\gamma}(C_n)\neq 0$, and $Prob_{\rho}(C_n') \neq
0$, which implies that $\alpha$ and $\beta$ need not vanish.  In
particular, if $\alpha \neq 0$, then PBC implies that $\tbe=0$ and one
cannot deduce anything about $\beta$. Thus, there is no contradiction
with the condition of PBC in taking $\alpha \neq 0$.

When GC is imposed, then $Prob_{\gamma}(C_n)= Prob_{\gamma'}(C_n) =
Prob(C_n)$ and similarly $Prob_{\rho}(C_n')= Prob_{\rho'}(C_n') =
Prob(C_n')$.  Note however, that even without imposing GC, WBC
requires specifically that $\alpha=0$ which then {\sl implies} that GC
is satisfied for the pairs of paths $(\gamma, \gamma')$ and $(\rho,
\rho')$.  Indeed, the differences between the two generalised Bell
causality conditions remain even after imposing GC, since for PBC
$Prob_{\gamma}(C_n)= Prob_{\gamma'}(C_n)=0$ does not imply
specifically that $\alpha=0$. 

Nevertheless, the resulting dynamics in both cases is EP, as we will
show in the appendix. This can be traced to the fact that if
$Prob(C_{n-1})=0$, the transitions from $C_{n-1}$ become irrelevant to
the dynamics.

\noindent {\bf Acknowledgements:} We would like to thank Graham
Brightwell, Jeremy Butterfield, Djamel Dou, David Rideout, Rafael
Sorkin and Madhavan Varadarajan for discussions.

\bibliography{../Bibliography/refs}

\providecommand{\href}[2]{#2}\begingroup\raggedright\begin{thebibliography}{1}

\bibitem{Rideout:1999ub}
D.~P. Rideout and R.~D. Sorkin, {\it A classical sequential growth dynamics for
  causal sets},  {\em Phys. Rev.} {\bf D61} (2000) 024002,
  [\href{http://xxx.lanl.gov/abs/gr-qc/9904062}{{\tt gr-qc/9904062}}].

\bibitem{Henson:2004wb}
J.~Henson, {\it Comparing causality principles},
  \href{http://xxx.lanl.gov/abs/quant-ph/0410051}{{\tt quant-ph/0410051}}.

\bibitem{Rideout:2005}
D.~P. Rideout and M.~Varadarajan, {\it 
A general solution for classical sequential
  growth dynamics of causal sets}, gr-qc/0504066. 

\bibitem{Brightwell:2002vw}
G.~Brightwell, H.~F. Dowker, R.~S. Garcia, J.~Henson, and R.~D. Sorkin, {\it
  `Observables' in causal set cosmology},  {\em Phys. Rev.} {\bf D67} (2003)
  084031, [\href{http://xxx.lanl.gov/abs/gr-qc/0210061}{{\tt gr-qc/0210061}}].

\bibitem{HawkingEllis}
S.~W. Hawking and G.~F.~R. Ellis, {\em The Large Scale Structure of
  Space-Time}.
\newblock Cambridge University Press, Cambridge (UK), 1973.

\bibitem{Seuss}
Dr.~Seuss, {\em Yertle the turtle and other stories}.
\newblock Random House, New York, 1958.

\end{thebibliography}\endgroup
\bibliographystyle{../Bibliography/JHEP}

\renewcommand{\theequation}{A-\arabic{equation}}
  \setcounter{equation}{0}  
  \section*{Appendix: PBC + GC $\Rightarrow$ EP}  

\noindent {\bf Lemma A-1:} (A) Let $C_n$ be an $n$-element causal set
with $Prob(C_n) \neq 0$.  Let $q_i \neq 0$ for $0\leq i \leq n-1$.
Then the gregarious transition $g$ from $C_n$ is given by the
$n$-antichain to $(n+1)$-antichain transition $q_n$. (B) Let
$C_{(a)n}$ be an $(n+|a|)$-element causal set with break at rank
$|a|$ with past $a$, such that $Prob(C_{(a)n}) \neq 0$. Let $q_{(a)i}
\neq 0$ for $0\leq i \leq (n(a)-1)$.  Then the $a$-relative gregarious
transition $g_{(a)}$ from $C_{(a)n}$ is given by the $a$-relative
$n$-antichain to $a$-relative $(n+1)$-antichain transition $q_{(a)n}$.

\noindent {\bf Proof:} Define an {\sl atomisation} of $C_n^{(0)}$ as
follows.  For any $C_n^{(0)}$ there exists a $k \leq n $ such that
$C_n^{(0)}$ can be grown from a $k$-antichain $C_k^{(0)}$ along some
path $\gamma$.  Let $C_{k+i}^{(0)}$ represent the $i$th element in
this growth, with $0 \leq i \leq n-k$. An atomisation of $C_n^{(0)}$
is then the set of $n$-element causets $C_n^{(i)} \equiv C_{n-i}^{(0)}
\sqcup \,$ $i$-antichain. Such an atomisation plays a crucial role in
the proof of Lemma 2 in \cite{Rideout:1999ub}.  
We will also need to consider the
set of causets labelled by $i, j$, $C_{(k+i)}^{(j)} \equiv
C_{(k+i-j)}^{(0)} \sqcup $ $j$-antichain, with $0 \leq i \leq n-k$ and
$0\leq j \leq i$.  The $C_{(k+i)}^{(i)}$ are then $(k+i)$-antichains.

Define the transitions
\begin{eqnarray}
\alpha_i^{(j)}:C_{(k+i)}^{(j)} &\rightarrow&  C_{(k+i+1)}^{(j)} \\
 \beta_i^{(j)}  : C_{(k+i)}^{(j)}& \rightarrow & C_{(k+i+1)}^{(j+1)},
\end{eqnarray}
so that $\beta_i^{(i)}= q_{(k+i)}$. Each $\beta_i^{(j)}$ is therefore
a gregarious transition with a spectator set $S_\beta =
C_{(k+i)}^{(j)}$. The decomposition $C_{(k+i)}^{(j)} =
C_{(k+i-j)}^{(0)} \sqcup$ $j$-antichain, then tells us that
$\alpha_i^{(j)}$ is a bold transition with a spectator set $S_\alpha
\supset j-antichain$,  if $j\neq 0$. Therefore,  the common spectator set
for $\alpha_i^{(j)}$ and $\beta_i^{(j)}$ includes the $j$-antichain.
The transitions without the $j$-antichain spectators are then
$\alpha_{i-j}^{(0)}$ and $\beta_{i-j}^{(0)}$, from $C_{k+i-j}^{(0)}$.
PBC then tells us that
 \begin{equation} \label{star}
\beta_i^{(j)} \alpha_{(0)}^{(i-j)}  =  \beta_{i-j}^{(0)}
\alpha_{i}^{(j)}.
\end{equation}
PBC and GC can be combined to give
\begin{equation} \label{ast}
\beta_i^{(j)} \alpha_{i-1}^{(j-1)}  =  \beta_{i}^{(j-1)}
\alpha_{i-1}^{(j-1)} \Rightarrow    \beta_i^{(j)} =
\beta_{i}^{(j-1)}, \quad  if  \quad \alpha_{i-1}^{(j-1)} \neq 0.
\end{equation}
Moreover, since  $Prob(C_n^{(0)}) \neq 0 $,
\begin{equation} \label{prob}
\alpha_i^{(0)} \neq 0, \, \, \forall \, \,   0\leq i \leq n-k,  \quad q_{l} \neq 0 \, \forall \, \, l < k.
\end{equation}

\noindent (A) We need to prove that $g= \beta_{n-k}^{(0)}=
\beta_{n-k}^{(n-k)}$ when $q_i \neq 0$, $ \forall \,\,i<n$.  If $n=k$
then we're done, since $g=q_k=\beta_0^{(0)}$. Let us therefore assume
that $n>k$. We use a proof by induction.  Let $j=i$ in (\ref{star}).
Since $\alpha_0^{(0)}$ and $\beta_0^{(0)}$ are non-zero,
$\alpha_i^{(i)} \neq 0$ iff $\beta_i^{(i)} \neq 0$. Since the latter
is non-vanishing for all $0\leq i<n-k$, $\alpha_i^{(i)} \neq 0 \, \,
\forall \, \, 0\leq i < n-k$. Putting $j=i$ in (\ref{ast}) then gives
$\beta_i^{(i)}=\beta_i^{(i-1)} = q_{k+i} \, \, \forall \, \, 0 \leq i
< n-k$. Putting $j=i-1$, and using these results we can deduce that
$\alpha_i^{(i-1)} \neq 0 \, \, \forall \, \, 0\leq i < n-k$ and
$\beta_i^{(i-1)}=\beta_i^{(i-2)} = q_{k+i} \, \, \forall \, \, 0 \leq
i < n-k$. Let us assume this to be true for $j=i-s$, $s<n-k$,
i.e. that $\beta_i^{(i-s)}\neq 0 \, \, \forall \, \, 0 \leq i <
n-k$. Then $\alpha_i^{(i-s)} \neq 0$ from (\ref{star}) and
$\beta_i^{(i-s-1)}=\beta_i^{(i-s)} \neq 0$. Thus, by induction we see
that $\beta_i^{(i-s)}=\beta_i^{(i)}=q_{n+i} \neq 0 \,\, \forall \, \,
0 \leq i < n-k , s \leq i$, and $\alpha_i^{(i-s)} \neq 0 \,\, \forall
\, \, 0 \leq i < n-k , s \leq i$.

Finally, putting $i=n-k$ in (\ref{ast}), we see that
$\beta_{n-k}^{(j)}=\beta_{n-k}^{(j-1)} \, \, \forall \, \, 0 \leq j
 \leq n-k$, since $\alpha_{n-k-1}^{(j-1)} \neq 0$. Thus,
  $g=\beta_{n-k}^{(0)}= \beta_{n-k}^{(n-k)}= q_n$.

We note that this final step can be replaced by the proof for Lemma 2
in \cite{Rideout:1999ub}, if we replace BC with PBC and use the fact that $
Prob(C_{n}^{(i)}) \neq 0 \, \, \forall \,\,0 \leq i < n-k$: starting
from the $(k+i)$-antichain, consider the growth $C_{k+i}^{(i)}
\rightarrow C_{k+i+1}^{(i)} \rightarrow \ldots \rightarrow
C_{n}^{(i)}$, with the transition probabilities $\alpha_{i+s}^{(i)}:
C_{k+i+s}^{(i)} \rightarrow C_{k+i+s+1}^{(i)}$. Since none of these
transitions vanish, $ Prob(C_{n}^{(i)}) \neq 0 \, \, \forall \,\,0
\leq i < n-k$.

\noindent (B) The above proof allows a simple generalisation to this
 case. Namely replace $\alpha_i^{(j)}, \beta_i^{(j)}$ with
 $\alpha_{(a)i}^{(j)}, \beta_{(a)i}^{(j)}$, i.e. transitions relative
 to the causet $a$, and $C_{k+i}^{(j)}$ with $C_{(a)k+i}^{(j)}$,
 i.e. causets with a break at rank $|a|$ and past $a$. \qed

\noindent PBC then tells us that \\
\noindent {\bf Claim A-2:} Let $q_n$ be the first antichain transition to
vanish, i.e. $q_l \neq 0, l< n$.  Then the only non-vanishing
transitions at stage $n$ are the timid transitions. More generally,
let  $q_{(a)n(a)}$ be the first $a$-relative antichain transition to
vanish, i.e. $q_{(a)l} \neq 0, l< n(a)$.  Then the only non-vanishing
transitions at stage $n+n(a)$ are timid transitions.

\noindent {\bf Proof:} Let $g$ be the gregarious transition from some
$C_n$,  $Prob(C_n) \neq 0$. From the above Lemma, $g=q_n=0$. If $\beta$
is a bold transition from $C_n$, then it shares a set of common
spectators $S$. Removing these spectators gives us the pair of
transitions $\tbe, \tg$, from some $C_m=C_n \backslash S, m< n$, where
$\tg=q_m$. Thus, $g \tbe =\tg \beta \implies \beta =0$.  The
generalisation is straightforward.  \qed

\noindent {\bf Claim A-3:} Let $q_n$ be the first antichain transition to
vanish.  Any causal set $\tc \in \tO$ which is admitted  has a break at
rank $n$.  Similarly, if $q_{(a)n(a)}$ is the first $a$-relative
antichain transition to vanish, then any causet $\tc \in \tO$
which is admitted and has a break at rank $|a|$ with past $a$ also has a break
at rank $n(a)$.

\noindent {\bf Proof:} Assume otherwise. Let $\tc(n)$ be the
$n$-subcauset of $\tc$ and $e_m$ the first element in $\tc$, $m>n$
such that $\tc(n)$ does not belong to the past of $e_m$. Consider a
relabeling $\tc \rightarrow \tc'$ so that $\tc'(n)=\tc(n)$ and $e_m
\rightarrow e_{n+1}'$. Then adding $e_{n+1}'$ at stage $n$ corresponds
to a bold transition at stage $n$ which must vanish. Hence
$Prob(\tc'(n+1)) =0$ and hence $\tc$ does not occur. This proof
extends simply to the case of $a$ relative causets. \qed

As in \cite{Rideout:2005} we see the resulting dynamics is characterised by
the occurrence of ``eras''.  The arguments of \cite{Rideout:1999ub}
can then be simply carried over to show that within each era, the
dynamics is GP.  In particular, one can use PBC to prove Lemma 3 of
\cite{Rideout:1999ub} {\it within} each GP era. The resulting dynamics
thus has the form of extended percolation as described.

\end{document}